# GOLF - NG spectrometer, a space prototype for studying the dynamics of the deep solar interior

Sylvaine Turck-Chièze, Pierre-Henri Carton, Jérome Ballot, Jean-Christophe Barrière, Philippe Daniel-Thomas, Alain Delbart, Daniel Desforges, Rafaël A. Garcia, Rémi Granelli, Savita Mathur, François Nunio, Yves Piret,
*DAPNIA, CE Saclay, CEA, 91191 Gif sur Yvette Cedex, France  cturck@cea.fr*

Pere. L. Pallé, Antonio. J. Jiménez, Sebastian J. Jiménez-Reyes,
*IAC, Calle Via Lactea s/n, la Laguna, Ténérife, Spain*

Jean Maurice Robillot,
*Observatoire de Bordeaux, Bordeaux 1, France*

Eric Fossat,
*Université de Nice, France*

Antonio. M. Eff-Darwich & Bernard Gelly,
*Themis, C./ Saturno n3 38205 La Laguna, Ténérife, Spain*

**Abstract**

The GOLF-NG (Global Oscillations at Low Frequency New Generation) instrument is devoted to the search for solar gravity and acoustic modes, and also chromospheric modes from space. This instrument which is a successor to GOLF/SOHO will contribute to improve our knowledge of the dynamics of the solar radiative zone. It is a 15 points resonant scattering spectrometer, working on the D1 sodium line. A ground-based prototype is under construction to validate the difficult issues. It will be installed at the Teide Observatory, on Tenerife in 2006 to analyse the separation of the effects of the magnetic turbulence of the line from the solar oscillations. We are prepared to put a space version of this instrument including a capability of identification of the modes, in orbit during the next decade. This instrument should be included in the ILWS program as it offers a key to the improvement of our knowledge of the solar core in combination with observations from SDO and PICARD. We hope to determine the core rotation and magnetic field, through precise measurements of oscillation mode frequency splittings. Understanding the magnetic field of the radiative zone is important for progress in the study of solar activity sources, an important player for the long-term Sun-Earth relationship.

1. Introduction

Helioseismic measurements have been continuously improved since more than 25 years. After observations from one site, networks (BiSON, IRIS and GONG) and nowadays space instruments (GOLF, MDI, VIRGO aboard SOHO) have revealed important characteristics of the deep solar interior. Traditionally, the American teams have concentrated their efforts on Doppler imagers while European teams have mainly dedicated their efforts to global oscillations not only for the Sun but also for other stars.
   The present proposal of a new instrument for the ILWS program corresponds to this second category. The SOHO satellite has been successful in demonstrating the power of the Doppler velocity technique to get very precise information down to the solar core in measuring mode velocities as low as 1mm/s. But the detection of gravity modes remains unclear. This fact limits our knowledge of the core dynamics. We extend with GOLF-NG our expertise of the sodium line obtained with the ground-based IRIS network and the spaceborne GOLF instrument in order to reach extremely small signals (maybe down to 0.1 mm/s) as could be the case for gravity mode velocities at the solar surface. We need to decrease the effect of the solar granulation which appears as the main source of noise for the GOLF instrument at frequencies below 1 mHz where stand gravity waves. We need also to limit the instrumental noise to detect better and more quickly the modes around 1 mHz and below and above 4.5 mHz where some signal has been observed (Garcia et al. 1998). With such an instrument, we hope to improve our knowledge of the core dynamics and of the solar atmosphere up to the chromospheric region in studying the properties of the line (Eibe et al. 2001).

## 2. Scientific status and scientific goals

SOHO has been rich in discoveries. One major advance is certainly to begin to observe the different processes which produce the solar activity, which we are convinced is not only a surface effect. We are more and more conscious that the solar interior plays a crucial role, so we need to introduce magnetic field in the structural stellar equations to interpret properly the observations of the internal dynamics (rotation profile, meridional circulation, solar cycle (s), maybe some subtle consequent effect on the solar structure …). For doing this tremendous improvement, we must accumulate observations. By chance, SOHO has opened this avenue with the MDI instrument (Scherrer et al. 1995) and the Doppler resonant spectrometer GOLF (Gabriel et al. 1995), which continue to represent two very promising techniques for the near future.

In the first period of SOHO observations, most of the highlights have been concentrated on the convection zone and on the transition with the radiative zone, called "the tachocline" without which the solar dynamo cannot be understood (Dikpati and Gilman 2001). Today, we cannot imagine understanding long term magnetic evolution without a good insight into the internal dynamics of the whole sun, including the deeper layers probed by global oscillations (Bertello et al. 2000, Garcia et al. 2001, 2004). The dynamics of the radiative zone is more difficult to reach than the dynamics of the convective zone because the acoustic modes are principally sensitive to the outer layers, but the instruments GOLF and MDI have extracted the low frequency acoustic modes which are less influenced by surface effects and sensitive to the deep interior structure. This advance leads today to an unprecedented accuracy on the sound speed profile in the solar core (Turck-Chièze et al. 2001a, Couvidat et al. 2003a) and on the rotation profile in the radiative zone, down to 0.2 $R_\odot$ (Couvidat et al. 2003b). Such progress has already provided important constraints for modelling. It is partly due to the very low instrumental noise of GOLF combined with the long and continuous observations. The rigid rotation between 0.4 to 0.7 $R_\odot$ has probably a magnetic origin. The presence of a magnetic field in the radiative zone blocks the diffusion of the tachocline shear layer into the radiative zone (Rüdiger & Kitchatinov 1997, Gough & McIntyre 1998, MacGregor & Charbonneau 1999). Consequently it may play some role in the solar cycle (s). In parallel, observations for practically two solar cycles of the global oscillations show the effect of the surface solar time variability on the modes (Jimenez-Reyes et al., 2003, 2004a,b).

Therefore new questions encourage new investigations and a real insight into the radiative dynamics down to the core requires several well identified detections of gravity modes. GOLF/SOHO has already improved the capability of detection by a factor 40 in comparison with ground observations, with a limit of detection in the mixed modes region of 2±1 mm/s (Gabriel et al. 2002, Turck-Chièze et al. 2004a). Gravity-mode candidates, identified with more than 98% confidence level after 8 years of visibility as not being pure noise (Turck-Chieze et al. 2004a,b) could be compatible with an increase of the rotation in the core and with a different rotation axis in this region than in the rest of the Sun. A relic of the solar system formation period is not excluded and suggests the presence of a strong magnetic field located in the central region which may be observed through hyperfine structure. It is on this scientific basis that GOLFNG has been designed, to contribute to answering the following questions:
- What is the rotation profile in the solar core?
- Is there a differential rotation near the limit of the nuclear core? (see Eff-Darwich 2004)
- Can we estimate the order of magnitude of the magnetic field in the radiative zone or more specifically in the nuclear burning core?
- Could we contribute to suggest other solar periodicities than the 22 year cycle?
- Can we put more constraints on the atmosphere up to 600 km and in the chromosphere.

These questions are fundamental questions because they are connected to the determination of the transport of energy inside the Sun (and other stars), which is presently insufficiently known. The solar models must be able to reproduce all the present seismic observations, including the rotation profile. One needs physics beyond the standard framework of stellar evolution (Mathis and Zahn 2004). For that, we need to understand the angular momentum transport in the radiative region where energy is emitted and not instantaneously transferred by radiation and we need to understand its variability if any.

These objectives are reachable if we improve the best techniques used in SOHO and also if we get simultaneous observations as we have done with GOLF and MDI for the detection of the low signals in the acoustic mode range. A more detailed review of the scientific challenges can be found in Turck-Chièze (2005a) and the roadmap for Europe dedicated to the future of the discipline is also discussed in Turck-Chièze (2005b). The present instrument will add some complementary information to the already programmed missions SDO/HMI and Picard which will be launched in 2008, in order to build a complete internal and dynamical view of the Sun.

## 3. The instrumental concept

A good insight into the dynamics of the radiative zone supposes a precise determination of the frequencies of the modes at low frequencies. A mode frequency accuracy of some nHz below 1mHz allows one to extract properly the rotational splittings, with an accuracy of a few %. This is very important for establishing any differential rotation profile in the radiative zone. It supposes also a correct extraction of very low velocities (of the order and below 1mm/s). To reach these objectives, one needs a very low instrumental noise and a reduction of the solar granulation or chromospheric noise.

The principle of the GOLF-NG instrument is to measure the Doppler shift of the D1 sodium Fraunhöfer solar line by a comparison with an absolute standard given by the sodium vapour cell, the heart of the experiment. A small portion of the absorption line provided by the resonance of the light in the vapour cell is split into its Zeeman components by means of a static longitudinal magnetic field whose strength varies along the longitudinal axis to explore different heights of the atmosphere. By changing the circular polarization of the incoming flux, it is possible to select 8 points on the right wing of the line or 8 points on the left wing, with one in common at the center of the line (see figure 1).

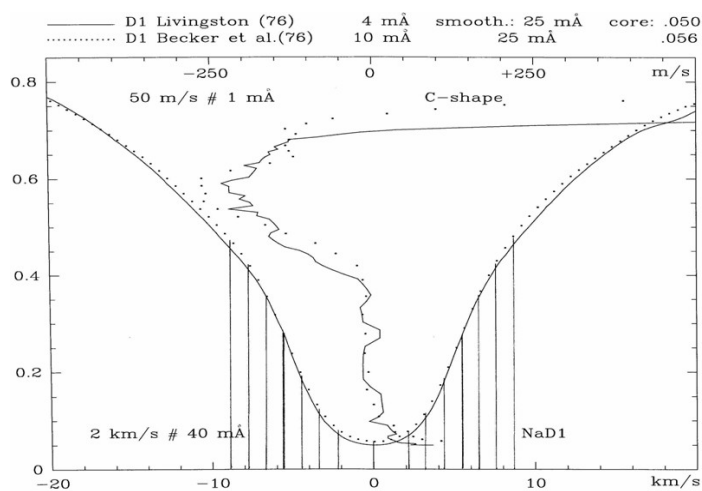

*Figure 1: The D1 sodium line as measured by Livingston or Becker in 1976 (private communication). Superimposed is an observed C-shape due to the solar granulation in the upper part of the line, The varying behaviour of this line corresponds to the upper abscissa (± 250 m/s). This line extends up to 500-600 km above the photosphere. The 15 channels of GOLF-NG are drawn and allow an excursion of ± 10 km/s (lower abscissa).*

The concept of a 15 points resonant spectrometer has been presented for the first time by Turck-Chièze et al. (2001b). It is an extension of the GOLF spectrometer (Gabriel et al. 1995) and of the 5 point spectrometer developed at Bordeaux for looking to the flow dynamics. Solar oscillations will be extracted from the bottom of the line to 50% of the continuum. Due to these 15 sampling points, the line is carefully determined and its magnetic deformations or palpitations will be followed in time. Espagnet et al. (1995) have shown that chromospheric dynamics dominates at an altitude of 500 km, while photospheric dynamics and granulation are the main contributors at lower altitude. They have shown that the Doppler velocity present different patterns at different heights. Playing with these different sources of noise is one important progress of the present instrument to disentangle the solar mode signals from solar noises. Doing so, it should be easier to extract the faint residual Doppler velocity signals induced by gravity modes. We have effectively demonstrated with the nominal 4 point GOLF instrument exploring 2 km/s (only operational during the first month of the mission), that we can improve the signal/noise in utilizing the signal at different heights or different sides of the line (Garcia et al. 2004) and suppress the non coherent part (70% of it on this small excursion). For GOLF-NG, about ± 10 km/s will be covered (see figure 1) so we hope to gain at least a factor 10 on the signal/ noise ratio in excluding a large part of the non coherent solar noise detected at different heights. Consequently we hope to explore part of the range of signal velocities suggested by theoretical estimates of these modes.

The small periodic signals are in fact lost in a velocity noise of about 1m/s and can be found through its temporal coherence in contrast with the non coherent noise in a long duration observation. The original measurement is a velocity signal of about 1 km/s which is due to the orbital motion of the satellite around the Sun and the gravitational signal. Figure 2 shows the power spectrum of the GOLF instrument and the different sources of noise.

The GOLF-NG instrument should also help to understand the different sources of the solar noise which are good indicators of the atmosphere. The high frequency part of the spectrum will be devoted to the search for chromospheric modes. As the statistics will be higher, the level of detection will be lower (see section 4). Moreover a simultaneous measurement of the continuum around the sodium line will help to disentangle intensity from velocity measurements.

The expected instrumental sensitivity will be of the order of $10^{-7}$ to reach properly the range of the gravity modes, it is an order of magnitude better than in the GOLF instrument. To reach this goal, some key-issues are already being checked with laboratory tests, it is the main reason to build a prototype and test different detectors.

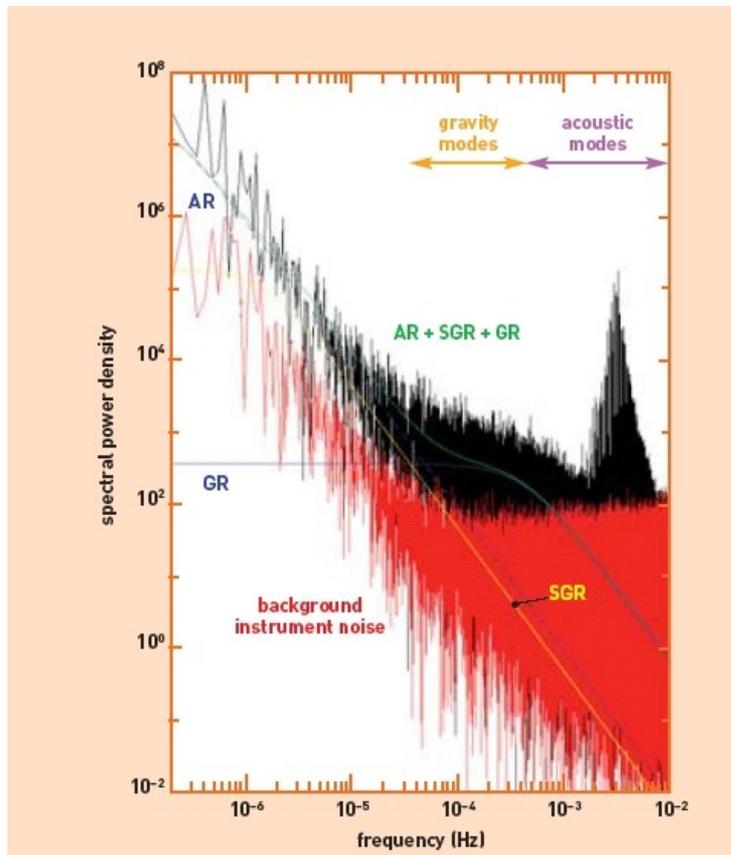

*Figure 2: Solar power spectrum density obtained with the GOLF instrument after two years of observations (Turck-Chièze et al. 2004a). The solar granulation noise dominates the region of gravity modes below 0.5 mHz. The instrumental noise was dominated by the statistical noise (in red or grey), at the beginning of the mission. The objectives of GOLF-NG are to reduce both by an order of magnitude.*

## 4. The instrument description

The principle of the instrument is summarized on Figure 3 and described here:
- The solar light passes through a 50 Å band pass filter which provides an UV and infrared protection and arrives on a first lens L1.
- A first magnet selector chooses between sun or white light for laboratory tests.
- A 40 cm multimode optical fibre guarantees the homogeneity of the sunlight and removes any incident polarisation.
- The optical system (L2 +L3) produces a rather parallel beam along the cell
- The second magnet selector allows the selection of the narrow filters: filter 5896 Å for the scientific case, filter 5910 Å centered on the continuum for checking the scattering beam in the cell (optional filter during flight, it may follow the time evolution of the cell without changing the cell temperature)
- The narrow filter at 5896 Å isolates a 5 Å band centred on the Na D1 line. It contains a large range of the continuum + the NaD1 line.
- The beam splitter cube separates the beam in two parts. One, (dominated in fact by the continuum) is directly sent to the detector box. This added information allows us to disentangle the intensity variation of the line from the velocity variation and to measure the ratio signal/continuum every 5s.
- The other solar one is circularly polarised right or left by *a liquid crystal retarder* before traversing the sodium vapour cell. Doing so, one selects the $\sigma+$ or $\sigma-$ Zeeman components, thereby, selecting blue (left) or red (right) wings.
- The cell is placed inside a varying permanent magnet. Its field topology guarantees a practically linear variation of the magnetic field between 0 and 12 kG. The resonant light scattered from the vapour cell maintained at a temperature around 200°C is finally extracted by 8 outputs along the cell. The 8 points correspond to values: 0, 2, 3, 4, 5, 6, 7, 8 kG. The sodium cell is equivalent to a very narrow band pass filter of 30 m Å (the intrinsic width of the absorption line).
So the line is carefully determined from the bottom up to 50% of the continuum.
- Four optical fibres bring the light flux to the detectors for each of these points. That means a total of 32-1=31 outputs, in fact 1 is missing to let some place for the stem of the cell.
- The mean required counting rates of the detectors are $10^8$ photons/s at mid altitude (instead $10^7$ for GOLF). The counting rate varies by a factor 10 according to the sampling position along the line. At the mean altitude each detector receives $2.5\ 10^7$ photons/s because four outputs are placed around the cell for each extraction position to include redundancies and the possibility of instrumental noise reduction.

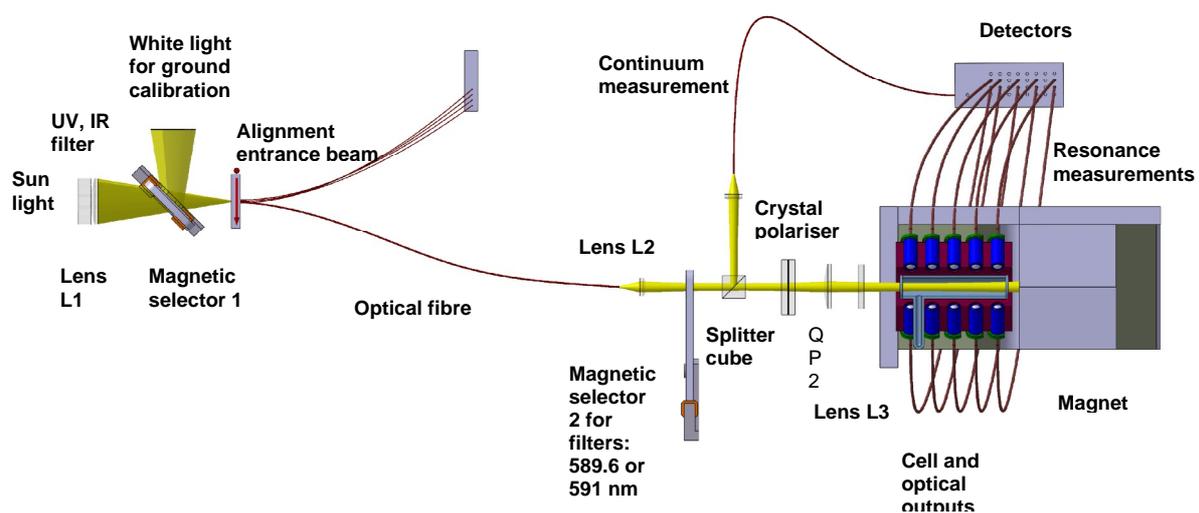

*Figure 3: Principle of the resonant spectrometer prototype GOLF-NG.*

## 5. The present status

The performance of the instrument is difficult to achieve. It is why we have built a prototype to study all the difficult aspects of the design. Therefore, this strategy allows a gain of time for the space phase A and a low cost as different solutions have been studied in advance. The prototype is under construction in our laboratories. It is an international French-Spanish collaboration (PI: S. Turck-Chièze, Project Manager: PH Carton, Thermal and Mechanical Manager: F. Nunio, Instrument Project Manager: JM Robillot, Scientific Performance Manager: R. A. Garcia, Spanish supervisor and Site Manager: P. Palle). This prototype is financially supported by CEA and CNES for the French part and IAC and Space Spanish Agency for the Spanish part. The preparation of this instrument takes advantage of the European expertise on networks and space GOLF instrument aboard SOHO. This collaboration will hopefully be extended to other countries for the space project called DynaMICS (Turck-Chièze et al. 2005b).

The already constructed prototype (see figure 4 showing subsystems) has improved performance in comparison with the instruments presently in the ground-based networks. It will join the Tenerife site in 2006 where sodium line studies can be pursued. Nevertheless the required signal/noise ratio at low frequencies can only be achieved in space. One may note that the excursion along the line, during one day on earth is of the same order as the excursion between 2 points: about 0.5 km/s instead only 1 or 2 km/s for the L1 conditions along the orbit in the case of GOLF every 6 months.

The prototype studies have already give solutions for the following important points:

(1) acquire a magnet (not too heavy) which produces a linear variation of the magnetic field to determine equidistant points along the cell on a small volume. The magnet has been realized by TE2M who has also delivered the GOLF magnet. It is a NdFeB magnet with polar elements in Fe-Co with 50% of Co and a saturated induction of 24.5 kG. The useful magnetic field along the cell extends from 0 to 8 kG.

(2) acquire a thermally equipped cell containing sodium. Its length is 6 cm. This element must be thermally stable in order to get a good and stable resonant factor. To study this important instrumental piece, a mechanical and thermal simulation has been done, then two subsystems studies have been performed: an optical study including all the different optical elements: filters, lens, and polarizer, and a thermal study including vacuum in the magnet to respect space conditions (magnet at 20-25°C and cell at 200°C with a stem at 175°C), thermal heaters and temperature sensors with an accuracy of 0.01°C. The preliminary phase studies (phase A) are finished, they have demonstrated mechanical and thermal solutions to achieve the expected performance, we are now checking them in the laboratory.

(3) limit the number of motors in the design for a long term use of the space instrument. This point has pushed two innovative studies:

- a magnetic selector with 2 positions which guarantees a nominal position (the normal one when the instrument is in operation), it is useful for the choice of the filters, it may be used for entrance polarisation if one needs in the space instrument.

- a liquid crystal retarder to avoid excess weight and mechanical problems. One changes the circular polarisation of the solar light every 5 s. For this important improvement of the GOLF-NG, we use a development undertaken by the Spanish group of IAC for the project IMAX aboard the balloon SUNRISE, the performance of the polarizer seems well adapted for our instrument.

- get appropriate detectors. The performance we need is extremely difficult to reach: very high counting rates during several years, + very low detector and electronic noises, small volume and uniformed response from one output to the other because we compare neighbour channels. The GOLF instrument was using Hamamatsu photon counting photomultipliers but these detectors have a too low quantum efficiency and a rather high dead time, moreover, the performance decreases with time due to the photocathode ageing. As we would like to count 4 times more per detector, and have 31 detectors instead 4 (in fact 2 for all the present scientific analyses), we are studying some other solutions. We have analysed the response of a matrix of 64 photodiodes S6494-64, provided by Hamamatsu, at reasonable temperature (6°C). The difficulty is to guarantee that the intrinsic noises (electronic and detector) are small fractions of the statistical noise, and that they evolve slowly during one solar cycle. Another study is under consideration, using an EEV CCD at very low temperature (-60-80°C). These detectors are used in the COROT project at higher temperature.

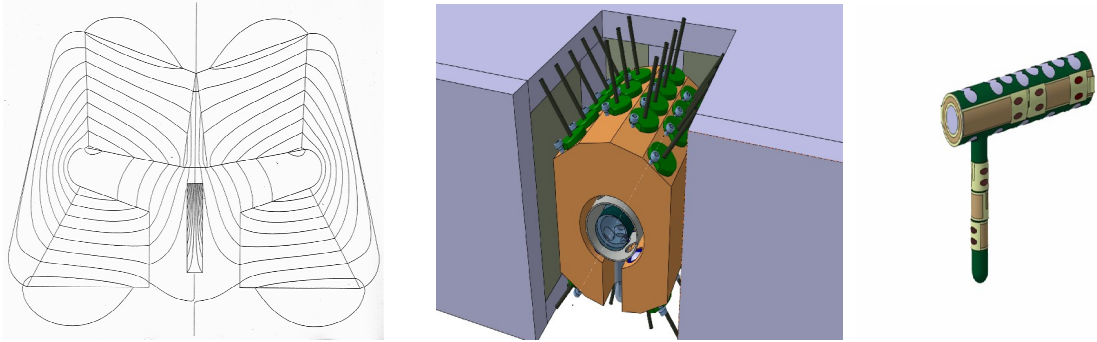

*Figure 4: Some subsystems of the GOLF-NG instrument. Left: the magnetic field lines of the magnet and the location of the cell. Middle: the cell inside the magnet with the 31 outputs, Right: the cell design equipped with the heaters.*

**6. The space strategy**

GOLF-NG represents the third generation of instruments which measure global Doppler velocity variability in the sodium line. It is dedicated to the detection of very low solar signal in space by a reduction of the different sources of noise. The sodium line allows this progress because it is a broad line with a smooth profile. This project uses the Zeeman splitting at different heights in the atmosphere thanks to a linearly varying permanent magnet. In the previously described design, GOLF-NG will measure only low degree modes (typically $\ell$ = 0, 1, 2, may be 3 or 4). For the space version, called DynAMICS (Dynamics And Magnetism of the Inner Core of the Sun), other improvements will be added: several masks at the entrance of the instrument will help in the identification of the degree of the modes up to $\ell$ = 5, moreover we are thinking about measuring global magnetic fields as was originally done in the GOLF design.

It will be extremely advantageous to launch this instrument as soon as possible to limit the perturbations of the maximum solar cycle which may limit the detection of very low signals (except if the mission observation is more than 6 years). It is crucial to launch it during observations done by other seismic space instruments to maximize the scientific return: the orbit of HMI gives little hope to progress on the dynamics of the radiative region of the Sun, if it observes alone. DynAMICS coupled with HMI will take advantage of the large range of degrees of HMI and of the noise reduction of this instrument. Together, we will put constraints on internal rotation and magnetic field from the core to the surface. It will be also a valuable companion of Picard which will benefit of the enhancement of the signal at the limb but which cannot reduce the solar noise.

Continuity with the SOHO effort is also important so the best strategy will be to launch such instrument as a demonstrator on a microsatellite and then as part of a huge world payload at the L1 Lagrange point for a long-term observation (see Turck-Chièze et al. 2005b). DynAMICS needs continuous observation, the Picard orbit (near polar low-Earth orbit, solar synchronous orbit) is appropriate and the SOHO orbit (around the L1 point) is perfect.

Preliminary characteristics of the space instrument have been established, they will be readjusted for specific mission profiles:
- dimension : 0.6 *0.3 * 0.3 $m^3$
- mass between 30-50 kg (magnet 15 kg)
- power 60 W (this estimate could be reduced)
- telemetry 30kbits/s
- same pointing accuracy as Picard or HMI/SDO.

This instrument was proposed as part of the SOLARIS project payload in the F2/F3 call of ESA in 1998. The selection of Solar Orbiter has not offered to Europe the opportunity to introduce a helioseismic mission dedicated to the deep solar interior, even Picard will play a role in this direction. It is crucial to pursue these measurements in Europe, in parallel to the SDO efforts because gravity modes are key observables in the understanding of the magnetism of the whole Sun and only several g-modes will be a strong constraint to simulate the whole magnetic story of the Sun. The roadmap proposed to ESA shows a strong need to improve our observation of the radiative region (Turck-Chièze et al. 2005a,b) during the coming solar cycle and shows that Europe is well prepared to do this important step.

## 7. Acknowledgement

This project is inspired by European efforts of the last twenty years. We would like to thank our scientific colleagues and engineers who have contributed to BiSON, IRIS and GOLF, We would like also to thank the space agencies of our countries, together with CEA for financial and dedicated contracts.

## 8. References

Bertello L. et al., Identification of solar acoustic modes of low angular degree and low radial order, Astrophys. J. Lett. 537, L 143-146, 2000.
Couvidat S., Turck-Chièze, S., Kosovichev, A. G., Solar Seismic Models and the Neutrino Predictions, Astrophys. J.. 599, 1434-1448, 2003a.
Couvidat S., et al., The rotation of the deep solar layers, Astrophys. J. 597, L77-79, 2003b.
Dikpati, M., Gilman, P. A. Flux-Transport Dynamos with alpha-Effect from Global Instability of Tachocline Differential Rotation: A Solution for Magnetic Parity Selection in the Sun, Astrophys. J. 559, 428-442, 2001.
Eff Darwich, A., PCA Inversions for the Rotation of the Solar Radiative Interior, SOHO 14 / GONG 2004, ESA SP-559, p.420, 2004.
Eibe M. T., Mein, P., Roudier, Th & Faurobert, M., Investigation of temperature and velocity fluctuations through the solar photosphere with the Na1 D lines, Astron. Astrophys. 371, 1128, 2001.
Espagnet O., Muller R., Roudier, Th., Mein, N., Mein, P., Penetration of the granulation into the photosphere: height dependence of intensity and velocity fluctuations, Astron. Astrophys. S. Ser., 109, 79, 1995.
Gabriel A. H. et al., Global Oscillations at Low Frequency from SOHO mission (GOLF), Sol. Phys.162, 61-69, 1995.
Gabriel A.H., et al., A search for solar g modes in the GOLF data, Astron. Astroph., 390, 1119-1131, 2002.
Garcia, R. et al., High-Frequency Peaks in the Power Spectrum of Solar Velocity Observations from the GOLF Experiment, Astroph. J. 504, L51, 1998.
Garcia, R.A., et al. Low-Degree Low-Order Solar p Modes As Seen By GOLF On board SOHO, Sol. Phys. 200, 361-379, 2001.
Garcia, R. A., et al., About the rotation of the solar radiative interior, Sol. Phys. 220, 269-285, 2004.
Garcia, R. A., et al., Helioseismology of the Blue and Red Wings of the NA Profile as Seen by GOLF, , in SOHO14, ESA-SP559, p 432, 2004b.
Jimenez-Reyes, S. J. et al., Excitation and Damping of Low-Degree Solar p-Modes during Activity Cycle 23: Analysis of GOLF and VIRGO Sun Photometer Data, Astrophys. J. 595, 446-457, 2003.
Gough D. & McIntyre, M. E., Inevitability of a magnetic field in the Sun's radiative interior, Nature, 394, 755, 1998.
Jimenez-Reyes, S. J. et al., Tracing the ``Acoustic'' Solar Cycle: A Direct Comparison of BiSON and GOLF Low-l p-Mode Variations, Astrophys. J. 604, 969-976, 2004a.
Jimenez-Reyes, S. J. et al., On the Spatial Dependence of Low-Degree Solar p-Mode Frequency Shifts from Full-Disk and Resolved-Sun Observations, Astrophys. J. 610, L65-L68, 2004b.
Mathis, S. & Zahn, J.P., Transport and mixing in the radiation zones of rotating stars: I Hydrodynamical processes, Astron. & Astroph. 425, 229-242, 2004.
Mac Gregor, K. B. & Charbonneau, P., Angular Momentum Transport in Magnetized Stellar Radiative Zones. IV. Ferraro's Theorem and the Solar Tachocline, Astrophys. J., 519, 911-917, 1999,
Rüdiger, G., Kitchatinov, L. L., The slender solar tachocline: a magnetic model, Astron. Nachr. 318, 273, 1997.
Scherrer, P. et al., The Solar Oscillations Investigation- Michelson Doppler Imager, Sol. Phys. 129-188.
Turck-Chièze S. et al., Solar Neutrino Emission Deduced from a Seismic Model, Astrophys. J. 555, 69-73, 2001a.
Turck-Chièze, S. et al. g-mode: a new generation of helioseismic instrument SOHO 10/GONG 2000 Workshop, ESA SP-464, 331, 2001b,
Turck-Chièze S., et al., Looking for Gravity-Mode Multiplets with the GOLF Experiment aboard SOHO, Astrophys. J.. 604, 455- 468, 2004a; and erratum, Astrophys. J. 608, 610, 2004a.
Turck-Chièze, S. et al., Gravity modes with a resonant scattering spectrometer, in SOHO14, ESA-SP 559, 85, 2004b.
Turck-Chièze, S., Solar Gravity modes: Present and Future, COSPAR04-A- 03949, in press, 2005a.

Turck-Chièze, S. et al., The magnetism of the solar interior for a complete MHD solar vision, ESLAB proceedings, ESA-SP, in press, 2005b.